\documentclass[12pt,preprint]{aastex}

\newcommand{\mdot}{M$_{\odot}$ yr$^{-1}$}
\newcommand{\ldot}{L$_{\odot}$}

\def\kmsMpc{\ifmmode {\rm\,km\,s^{-1}\,Mpc^{-1}}\else
    ${\rm\,km\,s^{-1}\,Mpc^{-1}}$\fi}

\shorttitle{Star Formation Rates for Starburst Galaxies}
\shortauthors{Sargsyan and Weedman}

\begin{document}

\title{Star Formation Rates for Starburst Galaxies from Ultraviolet, Infrared, and Radio Luminosities}

\author{Lusine A. Sargsyan\altaffilmark{1} and Daniel W. Weedman\altaffilmark{2}}
\altaffiltext{1}{Byurakan Astrophysical Observatory and Isaac Newton Institute of Chile Armenian Branch, 378433, Byurakan, Aragatzotn Province, Armenia; sarl11@yahoo.com}
\altaffiltext {2}{Astronomy Department, Cornell University, Ithaca,
NY 14853; dweedman@isc.astro.cornell.edu}

\begin{abstract}
We present a comparison of star formation rates (SFR) determined from mid-infrared 7.7 $\mu$m PAH luminosity [SFR(PAH)], from 1.4 GHz radio luminosity [SFR(radio)], and from far ultraviolet luminosity [SFR(UV)] for a sample of 287 starburst galaxies with z $<$ 0.5 having $Spitzer$ IRS observations.  The previously adopted relation log [SFR(PAH)] = log [$\nu$L$_{\nu}$(7.7 $\mu$m)] - 42.57$\pm$0.2, for SFR in \mdot~ and $\nu$L$_{\nu}$(7.7 $\mu$m) the luminosity at the peak of the 7.7 $\mu$m PAH feature in ergs s$^{-1}$, is found to agree with SFR(radio).  Comparing with SFR(UV) determined independently from ultraviolet observations of the same sources with the GALEX mission (not corrected for dust extinction), the median log [SFR(PAH)/SFR(UV)] = 1.67, indicating that only 2\% of the ultraviolet continuum typically escapes extinction by dust within a starburst.  This ratio SFR(PAH)/SFR(UV) depends on infrared luminosity, with form log [SFR(PAH)/SFR(UV)] = (0.53$\pm$0.05)log [$\nu$L$_{\nu}$(7.7 $\mu$m)] - 21.5$\pm$0.18, indicating that more luminous starbursts are also dustier.  Using our adopted relation between $\nu$L$_{\nu}$(7.7 $\mu$m) and L$_{ir}$, this becomes log [SFR(PAH)/SFR(UV)]= (0.53$\pm$0.05)log L$_{ir}$ - 4.11$\pm$0.18, for L$_{ir}$ in \ldot.  Only Blue Compact Dwarf galaxies show comparable or greater SFR(UV) compared to SFR(PAH).  We also find that the ratio SFR(PAH)/SFR(UV) is similar to that in infrared-selected starbursts for a sample of Markarian starburst galaxies originally selected using optical classification, which implies that there is no significant selection effect in SFR(PAH)/SFR(UV) using starburst galaxies discovered by $Spitzer$. These results indicate that SFRs determined with ultraviolet luminosities require dust corrections by a factor of $\sim$ 10 for typical local starbursts but this factor increases to $>$ 700 for the most luminous starbursts at z $\sim$ 2.5.  Application of this factor explains why the most luminous starbursts discovered by $Spitzer$ at z $\sim$ 2.5 are optically faint; with this amount of extinction, the optical magnitude of a starburst having f$_{\nu}$(7.7 $\mu$m) of 1 mJy should be $V$ $\sim$ 25.6.


\end{abstract}



\section{Introduction}

The evolution of star formation in the universe is described by determining the total luminosity density arising from young stars as a function of redshift \citep[e.g. ][]{mad98,haa00,cal07}.  Various studies have attempted to measure cosmic evolution of this star formation rate (SFR) by locating galaxies using ultraviolet, optical, and near-infrared techniques \citep[e.g. ][]{mar07,lef05,man07,wee08}.  Because star-forming galaxies are discovered in various ways, it is vital to compare various methods for measuring the SFR in a galaxy.  The galaxies which we are studying in the present paper have their observed multiwavelength spectral characteristics determined primarily by on-going, luminous star formation.  For this reason, we use the classification "starburst galaxies", or "starbursts".  The starbursts which we include have SFRs ranging from $\sim$ 0.1 \mdot to $\sim$ 1000 \mdot.

Initial determinations of the evolution of SFR in the universe were based on optical observations which reveal rest-frame ultraviolet luminosities of starbursts at high redshifts \citep{mad98}.  Uncertain corrections for extinction are the greatest limitation of such studies, and the significance of dust obscuration has become increasingly emphasized by the large populations of starbursts being discovered among the dusty sources found by the Spitzer Space Telescope ($Spitzer$).  These "dust obscured galaxies" \citep{dey08} are revealed in various surveys at 24 $\mu$m with the Multiband Imaging Photometer for $Spitzer$ (MIPS; Rieke et al. 2004) combined with 3 $\mu$m to 8 $\mu$m photometry with the $Spitzer$ Infrared Array Camera (IRAC; Fazio et al. 2004).  

Luminous, dusty galaxies discovered with MIPS and IRAC have been characterized by spectra with The Infrared Spectrograph on $Spitzer$ (IRS; Houck et al. 2004) to redshifts z $\sim$ 3.  The starburst sources have a mid-infrared spectrum dominated by emission from Polycyclic Aromatic Hydrocarbons (PAH) \citep[e.g. ][]{yan07,wee06,far08,pop08}.  

It is especially important to have methods for measuring SFR at high redshifts using various observations enabled by $Spitzer$ because the mid-infrared luminosity is often the only measure available for starbursts in these luminous, distant, and heavily extincted sources.  A technique to measure SFR in many mid-infrared sources using a consistent parameter is to use the luminosity of the PAH features, because these arise in the photodissociation region at the inner edge of the dusty molecular clouds surrounding the starburst \citep{pee04}.  

In the present paper, we use a measurement of PAH luminosity to determine the SFR as measured in the infrared [SFR(PAH)] in order to compare with independent measures of SFR that arise from ultraviolet continuum data [SFR(UV)] and radio continuum data [SFR(radio)].  This comparison is enabled by the large number of starburst galaxies which have now been observed by the $Spitzer$ IRS and also by the Galaxy Evolution Explorer (GALEX, Martin et al. 2005) and at 1.4 Ghz by the Very Large Array, primarily in the FIRST survey \citep{whi97}.  We also utilize and classify the available optical spectra from the Sloan Digital Sky Survey (SDSS, Gunn et al. 1998).

Our comparison of SFR(PAH), SFR(radio), and SFR(UV) has three primary objectives: 1. to test the validity of the infrared PAH calibration of SFR; 2. to derive an estimate of extinction for the ultraviolet luminosity by an empirical measurement comparing SFR as measured in the infrared and the ultraviolet; 3. to determine if infrared-discovered starbursts are similar in dust content to starbursts discovered with other techniques.   We make these comparisons using a total sample of 287 starburst galaxies with $Spitzer$ IRS spectra, described below. 

As a control sample for comparison to these starbursts chosen because they have infrared observations, we use the original list of starburst galaxies derived from the Markarian galaxy sample and classified as starbursts with optical spectroscopy \citep{bal83}.  This sample was selected without knowledge of infrared characteristics and was discovered as galaxies having unusually bright continua in the visible ultraviolet \citep{mar67}.  Infrared flux densities measured at 25 $\mu$m by the Infrared Astronomical Satellite (IRAS) are compared to GALEX fluxes for a sample of 76 Markarian galaxies as a test of whether there are different results for SFR(PAH)/SFR(UV) when using the optically-discovered Markarian sample compared to the infrared-discovered $Spitzer$ sample.

\section{Sample Selection and Observations}

\subsection{ $Spitzer$ IRS Observations}

A summary of 243 starbursts with IRS spectra and determinations of SFR(PAH) is given by \citet{wee08}.  For comparison with GALEX and radio data, we consider only the 49 galaxies in this sample having z $<$ 0.5 and not included by \citet{bra06}.  The redshift limit is chosen both to enable measurements at an accessible ultraviolet rest-frame wavelength using the GALEX data, and to have a sufficiently bright sample that most of the $Spitzer$ sources are detected with GALAX and VLA observations.  The Brandl et al. starbursts are omitted, because part of our objective is to test the calibration for SFR determined from these galaxies.  Also, they are all resolved sources with the IRS, so uncertain aperture corrections would have to be applied for comparison to GALEX and VLA data.  The 49 sources which we use from \citet{wee08} are listed in Table 1, identified by footnote. 

To increase substantially the sample of $Spitzer$ starbursts, we have extracted new IRS spectra from the archival data of two $Spitzer$ Legacy Programs.  The first is program 30742 (D. Schiminovich, P.I.).  This program was designed to obtain IRS observations for sources within the Lockman Hole Field of the Spitzer Wide Area Infrared Extragalactic survey (SWIRE, Lonsdale et al. 2003) that also have measured fluxes from GALEX.  This program contains 101 sources with IRS spectra.  The second Legacy Program is program 40539 (G. Helou, P.I.) designed to obtain a flux limited sample from various $Spitzer$ survey fields of sources with f$_{\nu}$(24  $\mu$m) $>$ 5 mJy.  This program contains 330 sources.  

We extracted IRS spectra of all 431 sources in the two programs described above to find those sources which satisfy our criterion for a "pure starburst".  As in previous IRS studies of starbursts \citep{bra06,hou07,wee08,wee09}, this criterion for selecting sources whose mid-infrared spectrum is dominated by luminosity from a starburst (without being contaminated by AGN luminosity) is  having equivalent width (EW) of the 6.2 $\mu$m PAH feature be EW(6.2  $\mu$m) $>$ 0.4 $\mu$m.  The origin of this criterion is explained further in section 3.1, below. 

All of our new spectral extractions were done with the SMART analysis package \citep{hig04}, beginning with the Basic Calibrated Data products of the $Spitzer$ flux calibration pipeline, using the most recent version of calibration applied to the sources (either v.15 or v.17).  Because our objective was only to measure the PAH luminosity, we extracted only the IRS short-low spectra, which extend to $\sim$ 14 $\mu$m.  From the spectral extractions, we measured the redshift as determined from the PAH features at 6.2 $\mu$m, 7.7 $\mu$m, and 11.3 $\mu$m; measured the f$_{\nu}$(7.7  $\mu$m); and measured the rest-frame equivalent width of the 6.2 $\mu$m feature, EW(6.2  $\mu$m).  Results are in Table 1.

For the 101 sources in program 30742, 75 sources are found to satisfy the criterion of EW(6.2  $\mu$m) $>$ 0.4 $\mu$m.  These 75 sources are in Table 1, identified by footnote.  All of these sources have SDSS spectra available.  All show optical spectra classified as "HII", because the emission lines are narrow and the H$\beta$ emission line is strong compared to the higher ionization [OIII] lines.  This confirms the clear and strong correlation between the optical classification as a starburst and the presence of strong PAH emission.

For the 330 sources in program 40539, 152 sources are found to satisfy the starburst criterion EW(6.2  $\mu$m) $>$ 0.4 $\mu$m.  These 152 sources are in Table 1, identified by footnote. SDSS spectra are available for 57 of these sources, and all show optical spectra classified as "HII".  

Finally, we include 11 blue compact dwarfs (BCDs) from \citet{wu06} for which we extracted spectra to measure f$_{\nu}$(7.7  $\mu$m).  Most of these sources do not satisfy the EW(6.2  $\mu$m) $>$ 0.4 criterion because PAH features are systematically weak in BCDs.  We include these to have a sample of low luminosity sources. Adding these BCDs yields the total of 287 starbursts given in Table 1. 

Uncertainties in the measurement f$_{\nu}$(7.7 $\mu$m) for these sources arise from uncertainties in the $Spitzer$ flux calibration, which is typically $\pm$ 5\%, and uncertainties in fitting the value of the PAH peak at 7.7 $\mu$m.  For sources as bright as those we include, the combination of these uncertainties leads to a typical observational uncertainty in the infrared flux and luminosity measures of $\pm$ 10\%.

\subsection{GALEX and VLA Observations}

The GALEX mission is an all-sky survey satellite which obtains images in the ultraviolet at wavelengths 134-179 nm (FUV) and 177-283 nm (NUV) \citep{mar05}\footnote{GALEX data used in this paper were those available in February, 2009, at the MultiMission Archive at Space Telescope Science Institute [http://archive.stsci.edu/]}.  Point source images have full width at half maximum (FWHM) of 4.2\arcsec~ for FUV and 5.3\arcsec~ for NUV \citep{mor05}.  For the $Spitzer$ IRS, spatial resolution is wavelength dependent, so the FWHM of an unresolved source at the observed wavelength of the 7.7 $\mu$m feature depends on source redshift and is 3\arcsec~ $\la$ FWHM $\la$ 5\arcsec.  GALEX surveys report a one sigma positional uncertainty of 0.5\arcsec~ \citep{ag05} and $Spitzer$ surveys have a one-sigma uncertainty of 0.4\arcsec~ \citep{fad06}. 


These image sizes mean that the spatial resolutions of GALEX and $Spitzer$ observations are well matched.  Any mismatch does not affect measured fluxes for point sources, although there may be aperture-dependent effects between reported fluxes within the observing apertures if sources are very extended compared to the spatial resolution.  To avoid such uncertainties, we do not include in Table 1 any sources which are well resolved by the IRS (which is one reason why the nearby starburst galaxies in \citet{bra06} are not included), so the comparison of fluxes for $Spitzer$ and GALEX should not be significantly affected by the slightly differing spatial resolutions, even for any sources that are partially resolved.

A more significant source of uncertainty is determining whether $Spitzer$ and GALEX have detected the same source.  Confusion would arise, for example, if there were a highly obscured starburst found by $Spitzer$ but a less obscured, nearby starburst seen by GALEX.  To minimize this uncertainty, we use a positional-coincidence criterion.  Of the 287 $Spitzer$ infrared sources in Table 1, 283 were included in GALEX observations.  Of these 283 sources, 272 infrared sources in Table 1 have GALEX identifications to within 3\arcsec, and the GALEX observations for these sources are tabulated.  Sources with positional differences $>$ 3\arcsec~ are not tabulated because the GALEX source is outside the half-power location of the $Spitzer$ observing aperture.

The differences in position between $Spitzer$ and GALEX are shown in Figure 1 for all sources compared to source flux densities. For the 272 sources with distance $<$ 3\arcsec, the median difference between GALEX and $Spitzer$ positions is only 0.64\arcsec.  This value is comparable to the quoted positional uncertainties, which gives confidence that the same starbursts are observed in both infrared and ultraviolet.  

The observed flux density at rest frame FUV wavelength of 153 nm [used for determining SFR(UV)] is determined by using the GALEX observed flux densities at FUV (effective wavelength 153 nm) and NUV (effective wavelength 227 nm) to fit a power law continuum between these wavelengths, and then interpolate to the flux density at observed wavelength (1+z)153 nm using the resulting power law index.  This is the value of f$_{\nu}$(FUV) listed in Table 1. 

For comparing infrared flux densities to radio flux densities, similar considerations for spatial resolution and positional coincidence apply.  Most of the radio detections are from the FIRST survey \citep{whi97}, and the remainder are from targeted observations with the VLA \citep{con03}.  These surveys give spatial resolution with FWHM $\sim$ 5\arcsec, so the same positional criterion (agreement to within 3\arcsec~) is used for radio and infrared sources as for UV and infrared sources.  Of the sources in Table 1, 38\% have radio detections.  The radio limit which is used for sources that are not detected at 1.4 GHz is the 1 mJy limit of the FIRST survey.  Radio observations are corrected to rest frame 1.4 GHz by assuming a power law spectrum of index $\alpha$ = -0.7.  Results for the radio luminosities are in Table 1.

\section{Discussion}

\subsection {SFR from PAH Luminosities}

Many previous studies have shown a well defined correlation between the presence of PAH features in the mid-infrared spectrum and the presence of a starburst as classified in other parts of the spectrum \citep[e.g. ][]{gen98,rig00,bra06}.  This correlation occurs because the PAH emission arises in the photodissociation region that lies between the HII region of a starburst and the surrounding molecular cloud within which the stars formed.  The exceptional uniformity among the PAH spectrum of starburst galaxies is illustrated in Figure 2.  This Figure shows the overplotted, normalized IRS spectra of all 25 sources having an optical SDSS classification as starburst which are included within a $Spitzer$ flux limited sample defined only by f$_{\nu}$(24 $\mu$m) $>$ 10 mJy \citep{wee08}.  

The PAH features arise from vibrational modes in the complex PAH molecules \citep[e.g. ][]{pee02}. Variations in the intensity and hardness of the radiation field within starbursts change the relative strengths of the PAH features, but these changes are not great in the integrated spectra of starbursts, as seen in Figure 2.  The 6.2 $\mu$m and 7.7 $\mu$m features have very similar relative strengths, for example, because both arise from the CC stretching mode vibration.  By contrast, the relative strength of the 11.3 $\mu$m feature has greater variation because it arises from the CH bending mode.  

$Spitzer$ IRS spectra have been used to measure the SFR based on PAH measures and empirical calibration between PAH luminosities and bolometric luminosities \citep[e.g. ][]{bra06,pop08,hou07,wee08}.  The latter two papers describe and utilize the measure of PAH luminosity $\nu$L$_{\nu}$(7.7 $\mu$m), where 
L$_{\nu}$(7.7 $\mu$m) is determined only from the flux density at the peak of the 7.7 $\mu$m feature.  

The motivation for this measure of PAH luminosity is illustrated by the spectra in Figure 2.  The PAH features are complex and blended.  Determining the total luminosity of any single feature requires deblending and adopting a level for the underlying continuum of dust emission.  While such sophisticated deblending can be undertaken with sufficient S/N \citep{smi07}, it is not possible to do accurately in sources of limited wavelength baseline or poor S/N.  Because of the objective to measure PAH strength in high redshift sources with poor S/N and limited rest frame wavelength coverage, \citet{hou07} and \citet{wee08} define the PAH luminosity simply as $\nu$L$_{\nu}$(7.7 $\mu$m). 

The measure $\nu$L$_{\nu}$(7.7 $\mu$m) is not a measure only of the PAH luminosity, as would be an integrated measurement of a deblended PAH feature.  The measure based on peak $\nu$L$_{\nu}$(7.7 $\mu$m) includes some contribution from the dust continuum underlying the PAH feature.  As illustrated in Figure 2, this contribution is negligible.  For the representative continuum fit illustrated, this continuum contribution is $<$ 10\% of the peak f$_{\nu}$(7.7 $\mu$m).  For our use of $\nu$L$_{\nu}$(7.7 $\mu$m) as a measure of SFR, separating the PAH contribution from the dust continuum contribution is irrelevant so long as the dust continuum is also produced by the starburst.  This is the case because the calibration of SFR(PAH) in \citet{hou07} is done empirically using the total $\nu$L$_{\nu}$(7.7 $\mu$m), which includes both PAH and dust contributions. 

Systematic error in this use of $\nu$L$_{\nu}$(7.7 $\mu$m) to measure SFR(PAH) will arise only if the dust continuum has a significant contribution from dust heated by an AGN.  In this case, the underlying dust continuum will be enhanced, and $\nu$L$_{\nu}$(7.7 $\mu$m) will be artificially increased compared to that for a pure starburst.  This would lead to an overestimate of SFR(PAH).  To avoid this uncertainty, we deal only with sources which have sufficiently strong PAH features that they are pure starbursts without an AGN contribution to the mid-infrared continuum.
As discussed in \citet{hou07} and \citet{wee08}, the criterion for such starburst-dominated sources is that the EW of the 6.2 $\mu$m PAH feature exceed 0.4 $\mu$m in the source rest frame.

This criterion arises empirically from EW(6.2 $\mu$m) of the starbursts in \citet{bra06} classified optically and from the starbursts in \citet{wee08} which have SDSS optical classification as starbursts.  This quantitative criterion is also verified by the synthetic spectra in \citet{wee08} which combine the prototype starburst NGC 7714 with the AGN Markarian 231 and show that the starburst fraction of $\nu$L$_{\nu}$(7.7 $\mu$m) exceeds 90\% when EW(6.2 $\mu$m) $>$ 0.4 $\mu$m.  The sources we utilize in the present paper are sufficiently bright that the EW(6.2 $\mu$m) can be reliably measured, and these are given in Table 1.  While we could, therefore, use the total integrated luminosity in this feature as the measure of SFR for the sources in Table 1, a major objective of our study is to allow comparison of these sources to the many fainter sources discovered by $Spitzer$ at higher redshifts.  For these faint sources, f$_{\nu}$(7.7 $\mu$m) is a much more reliable observational measurement than is the total integrated flux in the 6.2 $\mu$m feature.  

To restrict the present analysis to sources for which we can be confident are dominated by a starburst, we also use the criterion of EW(6.2 $\mu$m) $\ga$ 0.4 $\mu$m. The only exception are 11 BCDs which are included from \citet{wu06}; these systematically have weaker PAH features than more luminous starbursts, although the BCDs show no spectral evidence of an AGN contribution.  The EW criterion used for the remaining starbursts is confirmed by the fact that all of the sources in Table 1 which have SDSS spectra show spectra classified as "HII", with no indication of an AGN of any type. 
 
For sources with a mid-infrared spectrum arising only from a starburst, the luminosity $\nu$L$_{\nu}$(7.7 $\mu$m) relates empirically to the total infrared luminosity of the starburst, $L_{ir}$, according to log $L_{ir}$ = log [$\nu$L$_{\nu}$(7.7 $\mu$m)] + 0.78 \citep{hou07}.  This transformation has no dependence on luminosity, as is seen by combining the IRAS-derived $L_{ir}$ for the local, low luminosity starburst sample of \citet{bra06} with the high luminosity, high redshift submillimeter galaxies in \citet{pop08}.  Using the relation from \citet{ken98}, this result for $L_{ir}$ yields that log [SFR(PAH)] = log [$\nu$L$_{\nu}$(7.7 $\mu$m)] - 42.57$\pm$0.2, for $\nu$L$_{\nu}$(7.7 $\mu$m) in ergs s$^{-1}$ and SFR in \mdot. 



\subsection{SFR from PAH Luminosities Compared to SFR from Radio Luminosities}

The measurement of SFR(PAH) using $\nu$L$_{\nu}$(7.7 $\mu$m) has been used to measure the local SFR density of the universe \citep{hou07} and to measure the evolution of starbursts \citep{wee08}.  It was found, for example, that star formation in the most luminous starbursts evolves as log[SFR(PAH] = 2.1($\pm$0.3) + 2.5($\pm$0.3) log(1+z), up to z = 2.5 for SFR(PAH in \mdot. 
In the present paper, our primary objective is to compare SFR(PAH) with SFR(UV).  Both to verify the previous conclusions based on SFR(PAH) and to undertake comparison with SFR(UV), it is 
useful to test the validity of the relation described above for SFR(PAH).  

\citet{hou07} showed that this relation was consistent to within $\sim$ 10\% for measure of SFR as determined from the luminosities of the infrared [Ne II] and [Ne III] lines \citep{ho07}, which arise in the HII region surrounding the starburst.  Also, \citet{hou07} showed that the empirical calibration of $\nu$L$_{\nu}$(7.7$\mu$m) to SFR leads to a measure of the integrated SFR in the local universe that is consistent with that measured for starburst galaxies using different samples and techniques for measuring SFR from infrared continuum luminosity.  Comparison with SFR determined with radio observations can provide an additional test of SFR(PAH).

The demonstration of a well defined correlation between far-infrared flux measured by IRAS and 1.4 GHz flux density from various radio observations \citep{hel85} led to a calibration of SFR(radio) as measured by 1.4 GHz luminosity \citep{con92}.  (Radio luminosity from a starburst arises primarily from the non-thermal luminosity of supernova remnants, and the number of such remnants relates to the SFR of the starburst.) This relation is log [SFR(radio)] = log [$\nu$L$_{\nu}$(1.4 GHz)] - 37.07, for $\nu$L$_{\nu}$(1.4 GHz) in ergs s$^{-1}$ and SFR(radio) in \mdot.  

Fundamentally, the calibration of SFR(radio) and the calibration of SFR(PAH) based on $\nu$L$_{\nu}$(7.7 $\mu$m) both arise from the premise of \citet{ken98} who equates the total bolometric luminosity radiated by dust to the bolometric luminosity of the young stars in the starburst.  Nevertheless, the calibrations of SFR(radio) and SFR(PAH) arise through different routes and using different sources so provide an independent consistency check on one another. 

Star formation rates SFR(PAH) measured with $\nu$L$_{\nu}$(7.7 $\mu$m) and SFR(radio) measured with $\nu$L$_{\nu}$(1.4 GHz) are given in Table 1.  Of the 287 infrared sources in Table 1, only 108 have actual radio detections.  For the remainder, SFR(radio) is determined as an upper limit assuming f$_{\nu}$(1.4 GHz) $<$ 1 mJy.  For these sources, SFR(PAH)/SFR(radio) is a lower limit.  

The comparison of SFR(PAH)/SFR(radio) for the detected radio sources is shown in Figure 3A. Sources range over a factor of 10$^{5}$ in luminosity but only a factor of 10$^{3}$ without the BCDs. The median value of the ratio log [SFR(PAH)/SFR(radio)] = 0.0 for these detected sources.  This indicates precise agreement between SFR(PAH) and SFR(radio) for starbursts, and there is no statistically meaningful dependence on luminosity for the starbursts which are not BCDs.

Only the BCDs depart from the agreement between SFR(PAH) and SFR(radio).  This result arises primarily because of the known weakness of PAH features in some BCDs \citep{wu06,iz98}, so that SFR(PAH) based on $\nu$L$_{\nu}$(7.7 $\mu$m) underestimates the true SFR. 

The sources in Table 1 without radio detections are plotted as limits in Figure 3B, using 1 mJy as the limit for f$_{\nu}$(1.4 Ghz).  The limits on SFR(PAH)/SFR(radio) shown in Figure 3B    
are in the sense that the actual SFR(PAH)/SFR(radio) would be greater than the limit plotted.  The distribution of limits has a median slightly smaller than the median of 0.0 for detected sources, which is as expected for the limits.


\subsection{SFR Determined from PAH Luminosities Compared to SFR from GALEX Ultraviolet Luminosities}

The most direct measurement of SFR in a starburst is to observe the ultraviolet continuum luminosity from the young, hot stars in the starburst \citep[e.g. ][]{sam07,tre07}.  The primary difficulty in obtaining SFR(UV) in this way is in determining the fraction of the observed luminosity which has been absorbed by intervening dust.  Corrections for this dust extinction can be determined observationally by measuring the reddening of the spectrum, using the Balmer decrement for example, and then using an extinction law to relate reddening to total extinction \citep{tre07}.  Such corrections cannot account, however, for luminosity sources so extinguished that they make no contribution to the observed spectrum from which reddening corrections are determined.

The same dust that absorbs ultraviolet luminosity from these "buried" sources produces infrared luminosity.  By comparing the infrared luminosity from the buried starburst with the observed ultraviolet luminosity, a direct measurement is made of the fraction of intrinsic ultraviolet luminosity which is absorbed.  It is this comparison that we undertake by comparing SFR(UV) and SFR(PAH), using only the observed ultraviolet luminosity without applying any corrections for dust attenuation.  

SFR(UV) is determined  using the relation SFR(UV)= 1.08x10$^{-28}$L$_{\nu}$(FUV), for L$_{\nu}$(FUV) in ergs s$^{-1}$ Hz$^{-1}$\citep{sam07}.  This is the intrinsic value, without applying any extinction corrections to the ultraviolet. For convenient comparison to SFR(PAH) and SFR(radio) in Table 1, we convert this to log [SFR(UV)] = log $\nu$L$_{\nu}$(FUV) - 43.26, for $\nu$L$_{\nu}$(FUV) in ergs s$^{-1}$ and SFR in \mdot, using f$_{\nu}$(FUV) determined as described above in section 2.2 for the FUV rest frame wavelength of 153 nm.   

The relations for SFR(PAH) and SFR(UV) also describe the correspondence between intrinsic flux $\nu$f$_{\nu}$(7.7 $\mu$m) for the PAH feature and FUV magnitude which should arise from the same starburst.  If SFR(PAH) = SFR(UV), and there is no extinction of the ultraviolet, log f$_{\nu}$(FUV) = log f$_{\nu}$(7.7 $\mu$m) - 1.01.  When compared to an infrared source for which f$_{\nu}$(7.7 $\mu$m) is 1 mJy, the intrinsic FUV magnitude of the same starburst in the AB system would be 
18.5 (using the transformation AB = -2.5log f$_{\nu}$ - 48.60, for f$_{\nu}$ in ergs cm$^{2}$ s$^{-1}$ Hz$^{-1}$.)

The comparison SFR(PAH)/SFR(UV) for the sources in Table 1 is shown in Figure 4.  In contrast to the close agreement between SFR(PAH) and SFR(radio) shown in Figure 3, there is a very large offset between SFR(PAH) and SFR(UV) in Figure 4.  The median log [SFR(PAH)/SFR(UV] = 1.67.  If this entire offset is caused by extinction of the ultraviolet luminosity, the median result indicates that only 2\% of the ultraviolet luminosity escapes the starburst, or that SFR(UV) would be underestimated by a factor of 50 if extinction were not considered.  

The median value is determined primarily by the large number of sources with log $\nu$L$_{\nu}$(7.7 $\mu$m) $>$ 43.  There is, however, a significant luminosity dependence in the ratio SFR(PAH)/SFR(UV).  Fitting all of the points in Figure 4 yields the result: 

\begin{equation}
$$log [SFR(PAH)/SFR(UV)]= (0.53$\pm$0.05)log [$\nu$L$_{\nu}$(7.7 $\mu$m)] - 21.49$\pm$0.18, for $\nu$L$_{\nu}$(7.7 $\mu$m) in ergs s$^{-1}$.$$
\end{equation}

\noindent The quoted errors arise only from the fitting of the line in order to demonstrate the intrinsic dispersion of the ratio SFR(PAH)/SFR(UV) among different starbursts; uncertainties for individual points are not included in the quoted errors.   


The most luminous starbursts are those which have been found by $Spitzer$ at z $\sim$ 2.5.  The luminosities scale with redshift \citep{wee08}, with log [$\nu$L$_{\nu}$(7.7 $\mu$m)] = 44.63 + 2.5log (1+z).  At z = 2.5, the most luminous starburst from this relation has log $\nu$L$_{\nu}$(7.7 $\mu$m) = 46.  Equation (1) and Figure 4 indicate that such starbursts should have values of SFR(PAH)/SFR(UV) $>$ 700!  

This very large ultraviolet extinction is consistent with the observation that the population of luminous $Spitzer$ sources at high redshift are very faint when observed optically \citep{dey08}.  At z = 2.5, the rest frame FUV wavelengths correspond approximately to observed $V$ magnitude.  The brightest obscured infrared sources at z = 2.5 have f$_{\nu}$(7.7 $\mu$m) $\sim$ 2 mJy in the observed frame.  According to the relation given above between 7.7 $\mu$m flux and rest frame ultraviolet magnitude, such a source with a factor of 700 extinction would have optical $V$ mag. $\sim$ 24.9.  


As was also the case for SFR(PAH)/SFR(radio) in Figure 3, the low luminosity BCDs in Figure 4 show systematically smaller values of SFR(PAH)/SFR(UV).  This result indicates low ultraviolet extinction in BCDs combined with weak PAH features. Various explanations related to low metallicity have been suggested to explain the weakness of PAH features in BCDs \citep{wu06,iz98}.  An alternative explanation for weak PAH features based on the observation of little extinction is that the HII regions are not ionization bounded, implying that the starbursts in BCDs are not enveloped by the dusty molecular clouds which produce the photodissociation region where PAH features arise in more luminous starbursts.

The same calibration of SFR(PAH) which we use was used by \citet{hou07} to determine the SFR density in the local universe from a complete, flux limited sample of $Spitzer$ sources.  The result was 0.008 \mdot Mpc$^{-3}$.  The local SFR determined with GALEX results was measured as 0.003 \mdot Mpc$^{-3}$ without ultraviolet extinction corrections \citep{igl06}, and increases to 0.016 \mdot Mpc$^{-3}$ with an assumed extinction correction of a factor of $\sim$ 5.  Based on comparisons with other infrared-derived measures of the local SFR density, Houck et al. consider their result based on SFR(PAH) to be underestimated by perhaps a factor of two because of the small number of low luminosity sources in their sample.  This implies that SFR(PAH) and SFR(UV) are consistent when extinction by a factor of 5 is applied to the ultraviolet data.

Why is this extinction correction by a factor of 5 for the local SFR so different from the median factor of 50 in SFR(PAH)/SFR(UV) in Figure 4?  The answer is in the luminosity dependence for SFR(PAH)/SFR(UV).  The SFR for the local sample \citep{hou07} is dominated by galaxies with 10$^{9}$ $<$ $L_{ir}$ $<$ 10$^{10}$ \ldot.  At these luminosities in Figure 4, 4 $<$ SFR(PAH)/SFR(UV) $<$ 16, which is consistent with the local correction of $\sim$ 5 applied for ultraviolet extinction.  

This result indicates that an infrared-discovered sample of dusty starbursts may contain extinction that is somewhat greater than that found in ultraviolet-discovered galaxies, but this excess extinction is less than a factor of two compared to the extinction within ultraviolet-discovered galaxies.  This consistency is evidence that results we find for starburst galaxies discovered in the infrared with $Spitzer$ can be used to consider the nature of starburst galaxies in general.  As a further test of this conclusion, we now consider a more detailed comparison with dust extinction in ultraviolet-selected galaxies.

\subsection{Dust Corrections for SFR(UV)}

The relations used above to determine SFR(PAH) and SFR(UV) derive from independent assumptions and modeling regarding the stellar composition of starbursts.  It would not be surprising if these independent approaches led to different values for SFR in a starburst.  We have assumed in our interpretation, however, that both approaches should give the same result for SFR and that the only differences arise from dust attenuation of the ultraviolet luminosity.  In order to test this assumption, we now consider independent measures of dust attenuation to determine if our conclusions regarding the amount of extinction by dust are consistent with other estimates.

The result in equation (1) and the distribution of points in Figure 4 indicates that the measured values for SFR(PAH)/SFR(UV) without corrections to SFR(UV) for dust extinction scale with starburst luminosity, such that SFR(PAH)/SFR(UV) is dependent on (starburst luminosity)$^{0.53}$.  If the ratio SFR(PAH)/SFR(UV) differs only because of differing dust extinction for the observed SFR(UV), this indicates that more luminous starbursts are also more dusty, which is one of our most important results.  For example, the points in Figure 4 indicate that dust extinction is negligible in starbursts of log[$\nu$L$_{\nu}$(7.7 $\mu$m)] $\sim$ 41 (typical luminosity of BCDs) but that dust extinction can reach a factor of 1000 in the most luminous starbursts. 

Our conclusion depends on the interpretation that the difference between observed values of SFR(PAH) and SFR(UV) arises only because of dust extinction of some fraction of intrinsic ultraviolet luminosity from the starburst.  This interpretation can be checked independently of our result.  Estimates of the dust extinction corrections that should be applied for ultraviolet luminosity have previously been made using extensive collections of galaxy SED templates to compare $L_{ir}$ to the total observed ultraviolet luminosity, $L_{uv}$, from galaxies \citet{bua07}.  The ratio $L_{ir}$/$L_{uv}$ derived in this way is a measure of dust extinction which is equivalent to our measure SFR(PAH)/SFR(UV).  These previous results also indicate a luminosity dependence for the dust extinction \citep{bua07}. 

The fundamental principle in using $L_{ir}$/$L_{uv}$ to measure dust attentuation of ultraviolet luminosity is that all infrared luminosity from a starburst arises from re-radiation by dust of ultraviolet luminosity which has been absorbed.  The total luminosity radiated by the starburst is then measured as $L_{ir}$ + $L_{uv}$.   This sum is the same as the intrinsic ultraviolet luminosity which is produced, which we define as $L_{uv}$(starburst).  The fraction of $L_{uv}$(starburst) which emerges without dust attenuation is $L_{uv}$/$L_{uv}$(starburst), or $L_{uv}$/($L_{ir}$ + $L_{uv}$).    Equivalently for our measurements, $L_{uv}$/$L_{uv}$(starburst) = SFR(UV)/[SFR(UV) + SFR(PAH)].

In order to use the results in \citet{bua07} with our relation in equation (1), we convert log $\nu$L$_{\nu}$(7.7 $\mu$m) to $L_{ir}$ using the empirical relation for starbursts in \citet{hou07}, log L$_{ir}$ = log [$\nu$L$_{\nu}$(7.7 $\mu$m)] - 32.80$\pm$0.2 for L$_{ir}$ in \ldot~and $\nu$L$_{\nu}$(7.7 $\mu$m) in ergs s$^{-1}$.  With this transformation, equation (1) becomes 

\begin{equation}
$$log [SFR(PAH)/SFR(UV)]= 0.53 log L$_{ir}$ - 4.11, for L$_{ir}$ in \ldot.$$
\end{equation}

We use equation (2) to determine the fraction of emerging ultraviolet luminosity, $L_{uv}$/$L_{uv}$(starburst) = SFR(UV)/[SFR(UV) + SFR(PAH)], and show this in Figure 5 as a function of luminosity.  Our result is compared to the same parameter for ultraviolet-selected galaxies using $L_{uv}$/($L_{ir}$ + $L_{uv}$) in \citet{bua07}.  The slope of the luminosity dependence is very similar between the two results.  Our relation from equation (2) is offset, however, by about a factor of two lower values of $L_{uv}$/$L_{uv}$(starburst).  This offset implies a systematically higher extinction by a factor of two at all luminosities for our infrared-discovered sample compared to the extinction estimates from ultraviolet-selected sources. 

There could be various reasons for the offset.  Part could be produced if our equation for SFR(PAH) is incorrect compared to SFR(UV) and gives larger values for SFR.  We concluded in section 3.2, however, that there is no offset between SFR(PAH) and SFR(radio).  Another explanation is that the templates in \citet{bua07} are not purely starbursts, so ultraviolet luminosity from portions of a galaxy with less extinction may be included than is the case for the pure starbursts in the infrared sample.  Finally, part of the excess extinction we find could also result from our selection based on infrared sources, which preferentially selects the most dusty starbursts.  We make another test for this final selection effect in section 3.5, below.



\subsection{Control Sample of Markarian Galaxies}


All of the starbursts considered so far were selected because they have $Spitzer$ infrared measurements, and most were discovered in $Spitzer$ surveys at 24 $\mu$m.   A concern is that this selection might lead to a selection effect which favors SFR(PAH) compared to SFR(UV).  To test for such an effect, we utilize a sample of starbursts selected without regard to their infrared fluxes but selected instead because of unusually strong fluxes in the visible ultraviolet.  This sample derives from the Markarian survey for "galaxies with strong ultraviolet continuum" \citep{mar67}.  It is the sample of 76 Markarian starbursts which have GALEX identifications among the Markarian galaxies defined as starbursts by \citet{bal83} using optical spectra which show only an HII spectrum with no indication of an AGN. 

The optical colors and emission line properties of these Markarian sources are explained by the presence of a hot, young star population, and this list was the first compilation of a starburst galaxy sample following the original definition of this galaxy category \citep{wee81}.  This sample is a useful control sample because most of the Markarian starbursts also have mid-infrared measurements by IRAS and ultraviolet measurements with GALEX, although neither mission existed when the original sample was assembled. In Table 2, the GALEX and IRAS data for these Markarian galaxies are given.  For sources detected by GALEX but not by IRAS, the 3$\sigma$ upper limit for f$_{\nu}$(25 $\mu$m) is taken as 65 mJy.  


To adopt an infrared spectrum, we use starburst spectra such as those in Figure 2 which characterize the unbiased samples of $Spitzer$ sources described in \citet{wee09}.  For sources in that sample having 42 $<$ log [$\nu$L$_{\nu}$ (15 $\mu$m)] $<$ 44, covering the luminosity range of most Markarian galaxies in Table 2, the average spectrum has f$_{\nu}$(25 $\mu$m)/f$_{\nu}$(7.7 $\mu$m) = 2.6$\pm$1.2, which allows a transformation from f$_{\nu}$(25 $\mu$m) observed by IRAS to a value of f$_{\nu}$(7.7 $\mu$m).  This allows the calculation of SFR(PAH) for the Markarian sample using the same relations used above for sources in Table 1 having observed values of f$_{\nu}$(7.7 $\mu$m).  Results for SFRs are in Table 2 and the ratios SFR(PAH)/SFR(UV) are shown in Figure 6. 

Fitting all of the points and limits in Figure 6 yields the result: 

\begin{equation}
 $$log[SFR(PAH)/SFR(UV)]= (0.38$\pm$0.08)log $\nu$L$_{\nu}$(7.7 $\mu$m) - 15.27$\pm$0.25.$$                       
\end{equation}

As for the infrared-selected starbursts in equation (1) and Figure 4, the Markarian starbursts in Figure 6 also show a dependence of SFR(PAH)/SFR(UV) on luminosity, although the range of luminosities is less than for the infrared-selected starbursts in Figure 4.  Within the uncertainties of equations (1) and (3), the ratio SFR(PAH)/SFR(UV) has the same dependence on luminosity.  At the median luminosity for Markarian galaxies in Figure 6 of log $\nu$L$_{\nu}$(7.7 $\mu$m) = 43, the log [SFR(PAH)/SFR(UV)] = 1.07.  At this same luminosity, the infrared-selected starbursts in Figure 4 have from equation (1), log [SFR(PAH)/SFR(UV)] = 1.30.  

This result indicates moderately less extinction in the Markarian galaxies; 8.5\% of the ultraviolet luminosity escapes compared to only 5\% which escapes the infrared-selected galaxies.  These differences are small compared to the large extinctions which are observed in both samples.  This means that extreme extinction in the ultraviolet is characteristic of all starbursts regardless of selection technique, so that choosing starbursts from infrared observations does not lead to a sample which is more heavily extincted by dust than choosing starbursts based on optical spectroscopic classifications.  

This comparison with Markarian galaxies is additional confirmation that an infrared-selected sample of starbursts does not yield starbursts which are significantly dustier than other starbursts.

\section{Summary and Conclusions}

Star formation rates determined from infrared ($Spitzer$), ultraviolet (GALEX), and radio (VLA) indicators have been determined for 287 starburst galaxies with spectra from the $Spitzer$ Infrared Spectrograph.  We find that SFR(PAH) and SFR(radio) give the same values, but SFR(UV) is typically a factor of 50 less than SFR(PAH).  This is an empirical measure of extinction in the ultraviolet.  

The measure of starburst luminosity used for SFR(PAH) is $\nu$L$_{\nu}$(7.7 $\mu$m) for the peak luminosity of the 7.7 $\mu$m PAH feature.  The measure of starburst luminosity for SFR(UV) arises from the observed continuum flux density at rest frame 153 nm.  The ratio log SFR(PAH)/SFR(UV) has a median value of 1.67, with dependence on luminosity of form log [SFR(PAH)/SFR(UV)] = (0.53$\pm$0.05)log [$\nu$L$_{\nu}$(7.7 $\mu$m)] - 21.49$\pm$0.18, for $\nu$L$_{\nu}$(7.7 $\mu$m) in ergs s$^{-1}$.   Using the conversion we adopt between $\nu$L$_{\nu}$(7.7 $\mu$m) and $L_{ir}$ for starbursts, this becomes log [SFR(PAH)/SFR(UV)]= 0.53 log L$_{ir}$ - 4.11, for L$_{ir}$ in \ldot.  

These results indicate that SFRs determined with ultraviolet luminosities require extinction corrections by a factor of $\sim$ 10 for typical local starbursts, but this correction factor exceeds 700 for the most luminous starbursts at z $\sim$ 2.5. The trend of increasing extinction with increasing luminosity is similar to that reported previously for ultraviolet-selected galaxies, although our infrared-selected sample shows a systematically larger dust extinction by about a factor of two. 

The extreme ultraviolet extinction explains why the most luminous starbursts discovered by $Spitzer$ at z $\sim$ 2.5 are optically faint.  These sources have log $\nu$L$_{\nu}$(7.7 $\mu$m) = 46.  With the derived amount of extinction, the optical magnitude of a starburst at z $\sim$ 2.5 with a 1 mJy 7.7 $\mu$m feature should be $V$ $\sim$ 25.6.

SFR(PAH) and SFR(UV) are compared for a control sample of 76 Markarian starburst galaxies having ultraviolet fluxes in order to test for selection effects arising from our use of infrared-selected starbursts. The result indicates only moderately less extinction in the Markarian galaxies; 8.5\% of the ultraviolet luminosity escapes galaxies compared to only 5\% which escapes infrared-selected galaxies of the same luminosity log $\nu$L$_{\nu}$(7.7 $\mu$m) $\sim$ 43.  This comparison with Markarian galaxies is additional confirmation that an infrared-selected sample of starbursts does not yield starbursts which are significantly dustier than other starbursts.   

Overall results for SFR(PAH) indicate that infrared-discovered starbursts are not significantly dustier that starbursts of similar luminosity discovered in other ways.  This means that conclusions derived from $Spitzer$ samples of starbursts are generally applicable to understanding the evolution of star formation in the universe.

\acknowledgments

We thank D. Barry for help in data acquisition, and P. Hall, V. Leboutiellier, and J. Bernard-Salas for aid in improving our IRS spectral analysis with SMART.  We thank the anonymous referee for suggestions which improved our paper.  This work is based primarily on observations made with the
Spitzer Space Telescope, which is operated by the Jet Propulsion
Laboratory, California Institute of Technology under NASA contract
1407. Support for this work at Cornell University was provided by NASA through Contract
Number 1257184 issued by JPL/Caltech.  Support was also provided by the US Civilian Research and Development Foundation under grant ARP1-2849-YE-06. This research has made use of the NASA/IPAC Extragalactic Database (NED) which is operated by JPL/Caltech under contract with NASA.

\clearpage
\pagestyle{empty}

\begin{deluxetable}{lcccccccccccccc} 
\rotate
\tablecolumns{15}
\tabletypesize{\footnotesize}
\tablewidth{-3pc}
\tablecaption{Infrared, Ultraviolet, and Radio Star Formation Rates of $Spitzer$ Starburst Galaxies}
\tablehead{
\colhead{No.}& \colhead{Name}& \colhead{coordinates} & \colhead{R\tablenotemark{a}} & \colhead{z\tablenotemark{b}}& \colhead{EW(6.2)\tablenotemark{c}} & \colhead{f$_{\nu}$(7.7)\tablenotemark{d}}& \colhead{$\nu$L$_{\nu}$(7.7)} & \colhead{f$_{\nu}$(FUV)\tablenotemark{e}}& \colhead{L(FUV)}& \colhead{f(1.4 GHz)\tablenotemark{f}} & \colhead{L(radio)} &  \colhead{SFR\tablenotemark{g}} & \colhead{SFR\tablenotemark{g}} & \colhead{SFR\tablenotemark{g}} \\
\colhead{} & \colhead{} & \colhead{J2000 for 24 $\mu$m}  & \colhead{\arcsec} & \colhead{} & \colhead{  $\mu$m} & \colhead{mJy} & \colhead{ergs s$^{-1}$} & \colhead{$\mu$Jy} & \colhead{ergs s$^{-1}$}  & \colhead{mJy} & \colhead{ergs s$^{-1}$} & \colhead{PAH} & \colhead{UV} & \colhead{Radio}
}



\startdata
1 & Haro 11\tablenotemark{k} & 003652.5-333319 & 1.1 & 0.021 & 0.13 & 20.4 & 43.84 & 2969 & 43.72 & 27.2 & 38.55 & 1.27 & 0.46 & 1.48 \\
2 & SST24\tablenotemark{j} & 021503.53-042422.0 & 1.5 & 0.141 & 0.60 & 3.2 & 43.76 & 1.8 & 42.2 & $<$1 & $<$38.81 & 1.19 & -1.06 & $<$1.74 \\
3 & SST24\tablenotemark{j} & 021557.13-033728.3 & 0.9 & 0.034 & 0.49 & 7.3 & 42.85 & 53.2 & 42.42 & $<$1 & $<$37.55 & 0.28 & -0.84 & $<$0.48 \\
4 & SST24\tablenotemark{j} & 021649.72-042554.3 & 0.3 & 0.141 & 0.57 & 6.1 & 44.03 & 52.5 & 43.67 & $<$1 & $<$38.81 & 1.46 & 0.41 & $<$1.74 \\
5 & SST24\tablenotemark{j} & 021743.85-051751.7 & 1.6 & 0.032 & 0.65 & 21 & 43.26 & 201.5 & 42.94 & 1.45 & 37.66 & 0.69 & -0.32 & 0.59 \\
6 & SST24\tablenotemark{j} & 021754.89-035825.8 & 0.4 & 0.226 & 0.55 & 10.6 & 44.70 & 38.3 & 43.96 & 1.41 & 39.38 & 2.13 & 0.70 & 2.31 \\
7 & SST24\tablenotemark{j} & 021849.78-052158.4 & 0.4 & 0.292 & 0.71 & 6.2 & 44.70 & 3.5 & 43.16 & 0.44 & 39.11 & 2.13 & -0.10 & 2.04 \\
8 & SST24\tablenotemark{j} & 021912.72-050541.4 & 0.3 & 0.195 & 0.88 & 2.6 & 43.95 & 38.5 & 43.83 & 0.4 & 38.71 & 1.38 & 0.57 & 1.64 \\
\enddata
\end{deluxetable}

\begin{deluxetable}{lccccccccc} 
\rotate
\tablecolumns{10}
\tabletypesize{\footnotesize}
\tablewidth{0pc}
\tablecaption{Infrared and Ultraviolet Star Formation Rates for Markarian Starburst Galaxies}
\tablehead{
\colhead{}&\colhead{Name}& \colhead{coordinates}& \colhead{z\tablenotemark{a}} & \colhead{f$_{\nu}$(25 $\mu$m)\tablenotemark{b}} & \colhead{f$_{\nu}$(FUV)\tablenotemark{c}} &\colhead{$\nu$L$_{\nu}$(FUV)} &\colhead{log[$\nu$L$_{\nu}$(7.7 $\mu$m)]\tablenotemark{d}} &\colhead{SFR(PAH)\tablenotemark{e}}& \colhead{SFR(UV)}\\
\colhead{}&\colhead{}& \colhead{J2000}& \colhead{} & \colhead{mJy} & \colhead{$\mu$Jy} & \colhead{ergs s$^{-1}$ Hz$^{-1}$} & \colhead{ergs s$^{-1}$}& \colhead{log\mdot }& \colhead{log\mdot}
}



\startdata
1 & Mrk 13 & 075614.20+601815.0 & 0.0046 & $<$ 65 & 476.9 & 41.62 & $<$41.64 & $<$-0.93 & -1.64 \\%
2 & Mrk 14 & 081059.30+724740.0 & 0.0081 & 151 & 883.5 & 42.39 & 42.51 & -0.06 & -0.87 \\
3 & Mrk 21 & 094830.80+575815.0 & 0.0281 & 109 & 304.5 & 43.01 & 43.45 & 0.88 & -0.25 \\
4 & Mrk 25 & 100351.80+592610.0 & 0.0100 & 251 & 817.4 & 42.53 & 42.91 & 0.34 & -0.73 \\
5 & Mrk 31 & 101942.80+572524.0 & 0.0260 & $<$ 65 & 135.4 & 42.59 & $<$43.16 & $<$0.59 & -0.67 \\
6 & Mrk 44 & 120324.50+384848.0 & 0.0227 & $<$ 65 & 122 & 42.43 & $<$43.04 & $<$0.47 & -0.83 \\%
7 & Mrk 52 & 122542.80+003422.0 & 0.0071 & 1050 & 2311.2 & 42.69 & 43.24 & 0.67 & -0.57 \\
8 & Mrk 87 & 082140.30+735918.0 & 0.0094 & 118 & 187.2 & 41.84 & 42.53 & -0.04 & -1.42 \\
\enddata
\tablenotetext{a}{Optical redshift from \citet{bal83}.}
\tablenotetext{b}{Flux density f$_{\nu}$(25 $\mu$m) from IRAS Faint Source Catalog.}
\tablenotetext{c}{Flux density at rest frame 153 nm interpolated with a power law between flux densities in the GALEX catalog measured with the FUV filter (134-179 nm) and the NUV filter (177-283 nm).}
\tablenotetext{d}{$\nu$L$_{\nu}$(7.7 $\mu$m) derived by transforming f$_{\nu}$(25 $\mu$m) to f$_{\nu}$(7.7 $\mu$m) using average infrared spectra for sources with 42 $<$ log [$\nu$L$_{\nu}$ (15 $\mu$m)] $<$ 44 in \citet{wee09}, for which f$_{\nu}$(25 $\mu$m)/f$_{\nu}$(7.7 $\mu$m) = 2.6$\pm$1.2.}
\tablenotetext{e} {Star formation rates log [SFR(PAH)] = log [$\nu$L$_{\nu}$(7.7 $\mu$m)] - 42.57 for $\nu$L$_{\nu}$(7.7 $\mu$m) the luminosity at the peak of the 7.7 $\mu$m feature in ergs s$^{-1}$ and SFR in \mdot; log [SFR(UV)] = log [$\nu$L$_{\nu}$(FUV)] - 43.26 for $\nu$L$_{\nu}$(FUV) in ergs s$^{-1}$ and SFR in \mdot (log [$\nu$L$_{\nu}$(\ldot)] = log [$\nu$L$_{\nu}$(ergs s$^{-1}$)] - 33.59).  Luminosity calculations use luminosity distances from E.L. Wright, http://www.astro.ucla.edu/~wright/CosmoCalc.html, for H$_0$ = 71 \kmsMpc, $\Omega_{M}$=0.27 and $\Omega_{\Lambda}$=0.73.}
\end{deluxetable}

\clearpage
\pagestyle{plaintop}
%
%
%

\begin{figure}
\figurenum{1}
\includegraphics[scale=0.9]{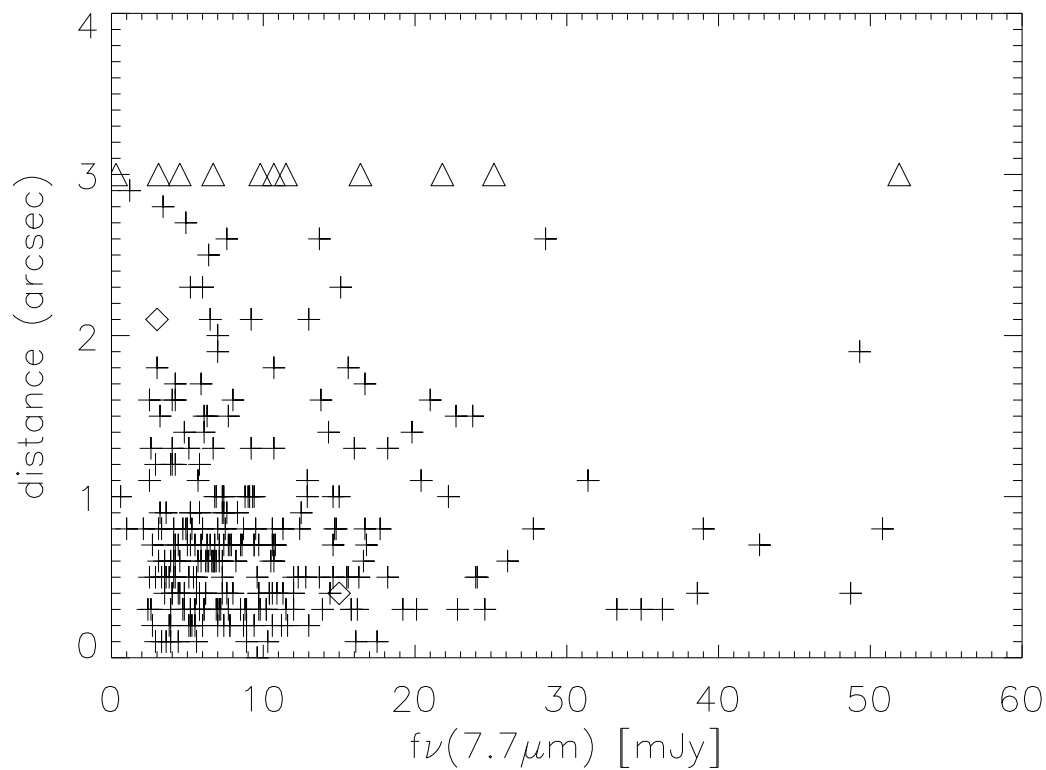}
\caption{Coordinate differences between GALEX and $Spitzer$ positions for sources in Table 1, shown as function of f$_{\nu}$(7.7$\mu$m) to illustrate that measured positional differences do not depend on brightness.  Crosses are measured differences; triangles are sources with differences $>$ 3\arcsec; diamonds are BCD sources without PAH features.  Median difference is 0.64\arcsec.   }

\end{figure}

\begin{figure}
\figurenum{2}
\includegraphics[scale=0.9]{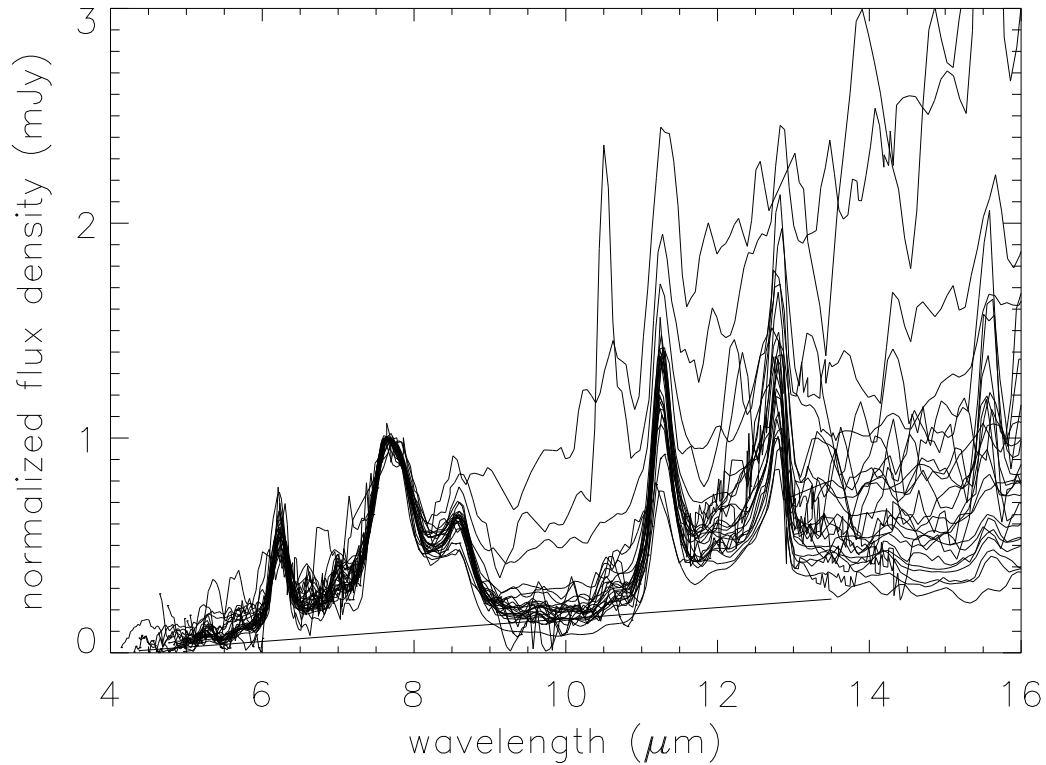}
\caption{Overplots of $Spitzer$ IRS spectra for all 25 sources in the $Spitzer$ 10 mJy sample of \citet{wee08} which have SDSS classifications as starburst (HII for SDSS classification, instead of any AGN classification).  All spectra are normalized to f$_{\nu}$(7.7$\mu$m) = 1 mJy; uniformity of the PAH features among starbursts is illustrated by the small scatter in PAH strengths. Primary spectroscopic differences are among the slopes of the dust continuum underlying PAH features; representative fit to dust continuum shown by diagonal solid line. Strong PAH emission feature at 7.7 $\mu$m is feature used in text to measure SFR(PAH) using $\nu$L$_{\nu}$(7.7 $\mu$m). }

\end{figure}

\begin{figure}
\figurenum{3A}
\includegraphics[scale=0.9]{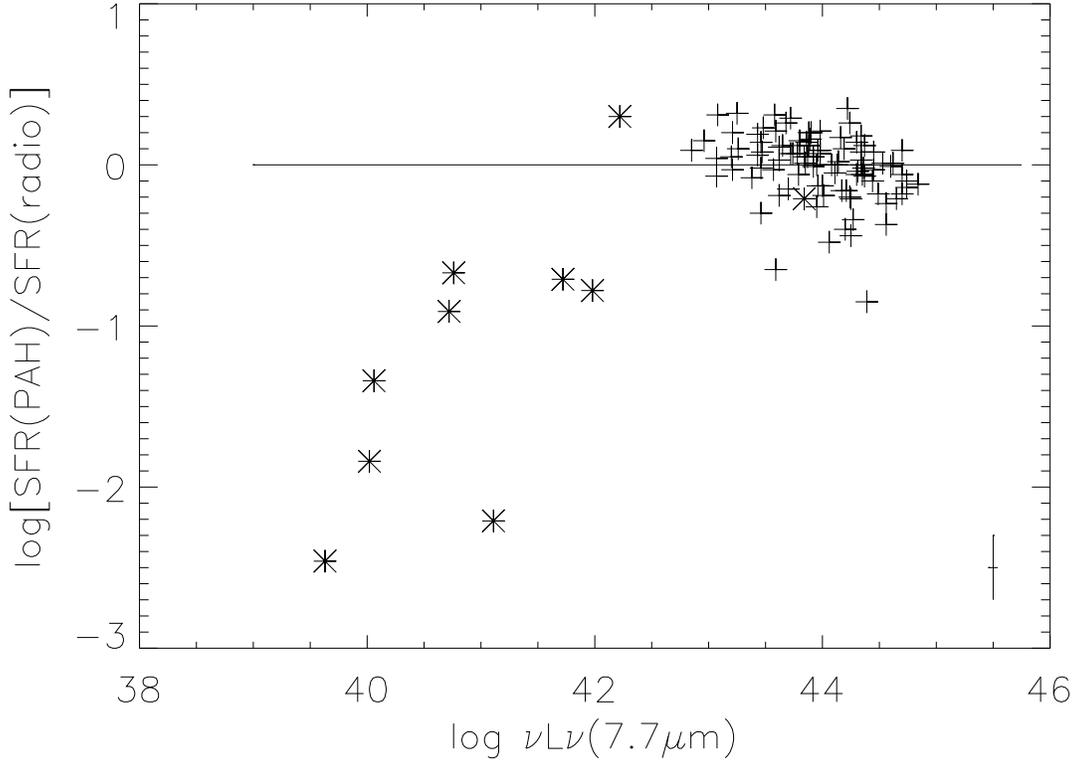}
\caption{Comparison of SFR(radio) measured from radio 1.4 GHz continuum luminosity to SFR(PAH) from 7.7 $\mu$m PAH luminosity for all sources with radio detections in Table 1 (asterisks are  BCDs).  Horizontal line is the median value of 0.0, indicating no offset in SFR(PAH) compared to SFR(radio).  $\nu$L$_{\nu}$(7.7 $\mu$m) is the luminosity at the peak of the 7.7 $\mu$m feature in ergs s$^{-1}$ (log [$\nu$L$_{\nu}$(7.7 $\mu$m)(\ldot)] = log [$\nu$L$_{\nu}$(7.7 $\mu$m)(ergs s$^{-1}$)] - 33.59).  SFR(radio) determined from log [SFR(radio)] = log [$\nu$L$_{\nu}$(1.4 Ghz)] - 37.07 for $\nu$L$_{\nu}$(1.4 Ghz) in ergs s$^{-1}$ and SFR(radio) in \mdot.  SFR(PAH) determined from log [SFR(PAH)] = log [$\nu$L$_{\nu}$(7.7 $\mu$m)] - 42.57.  Error cross in lower right hand corner shows 1$\sigma$ uncertainties in both directions; uncertainty for SFR(PAH) arises from uncertainty in determining transformation from $\nu$L$_{\nu}$(7.7 $\mu$m) to the total L$_{ir}$ which determines SFR; uncertainty in luminosity arises from $\sim$ 10\% uncertainty in measurement of f$_{\nu}$(7.7 $\mu$m). (Using empirically determined transformation between $\nu$L$_{\nu}$(7.7 $\mu$m) and total infrared luminosity L$_{ir}$, log L$_{ir}$ = log $\nu$L$_{\nu}$(7.7 $\mu$m) - 32.80$\pm$0.2 for L$_{ir}$ in \ldot~and $\nu$L$_{\nu}$(7.7 $\mu$m) in ergs s$^{-1}$.)}

\end{figure}

\begin{figure}
\figurenum{3B}
\includegraphics[scale=0.9]{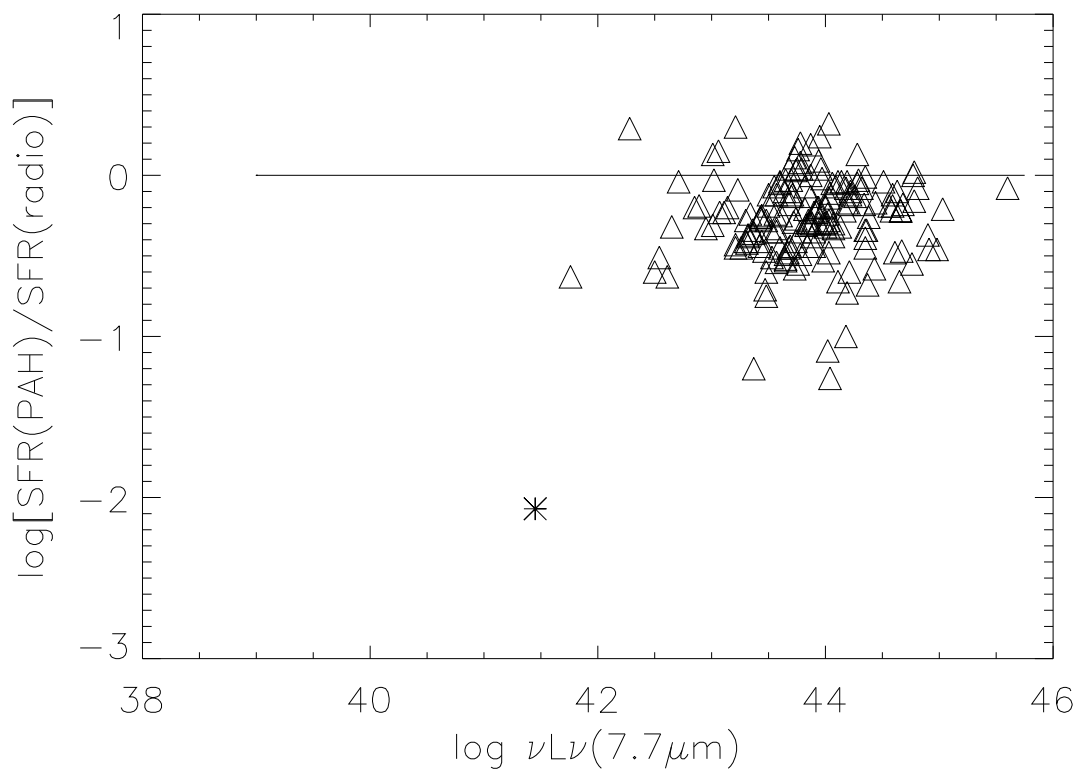}
\caption{Comparison of limits on SFR(radio) measured from radio 1.4 GHz continuum luminosity to SFR(PAH) from 7.7 $\mu$m PAH luminosity for all sources without radio detections in Table 1 using 1 mJy as the limit for f$_{\nu}$(1.4 Ghz) (asterisk is a BCD).  Limit is in the sense that SFR(PAH)/SFR(radio) is actually greater than the limit plotted.  Horizontal line is the median of 0.0 for the detected sources in Figure 3A; as expected, the median for limits is less than 0.0. }

\end{figure}

\begin{figure}
\figurenum{4}
\includegraphics[scale=0.9]{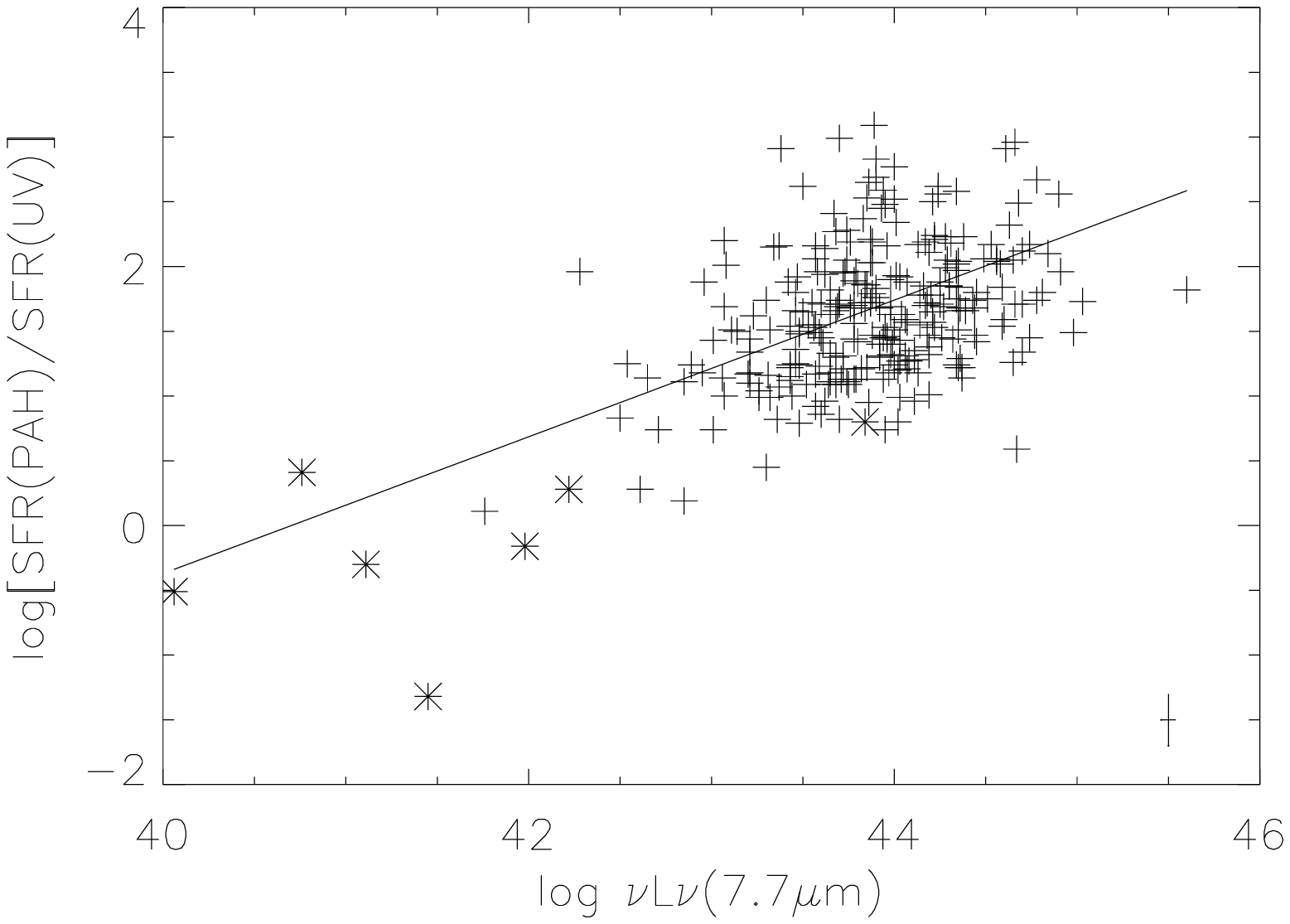}
\caption{Comparison of SFR(UV) measured from ultraviolet continuum luminosity to SFR(PAH) from infrared PAH luminosity for sources in Table 1 (asterisks are BCDs).  Ultraviolet SFR determined from log [SFR(UV)] = log [$\nu$L$_{\nu}$(FUV)] - 43.26 for $\nu$L$_{\nu}$(FUV) in ergs s$^{-1}$ and SFR(UV) in \mdot.  SFR(PAH) determined from log [SFR(PAH)] = log [$\nu$L$_{\nu}$(7.7 $\mu$m)] - 42.57.  SFR(PAH) exceeds SFR(UV), as would expected for extinction by dust of the ultraviolet continuum; median value of ratio log [SFR(PAH)/SFR(UV)] = 1.67, indicating that only 2\% of the ultraviolet continuum typically escapes a starburst; ratio increases with infrared PAH luminosity according to the linear fit which is shown, of form log [SFR(PAH)/SFR(UV)]= (0.53$\pm$0.05)log [$\nu$L$_{\nu}$(7.7 $\mu$m)] - 21.49$\pm$0.18. Error cross in lower right hand corner shows 1$\sigma$ uncertainties in both directions; uncertainty for SFR(PAH) arises from uncertainty in determining transformation from $\nu$L$_{\nu}$(7.7 $\mu$m) to the total L$_{ir}$ which determines SFR; uncertainty in luminosity arises from $\sim$ 10\% uncertainty in measurement of f$_{\nu}$(7.7 $\mu$m). (Using empirically determined transformation between $\nu$L$_{\nu}$(7.7 $\mu$m) and total infrared luminosity L$_{ir}$, log L$_{ir}$ = log $\nu$L$_{\nu}$(7.7 $\mu$m) - 32.80$\pm$0.2 for L$_{ir}$ in \ldot~and $\nu$L$_{\nu}$(7.7 $\mu$m) in ergs s$^{-1}$.) }


\end{figure}

\begin{figure}
\figurenum{5}
\includegraphics[scale=0.9]{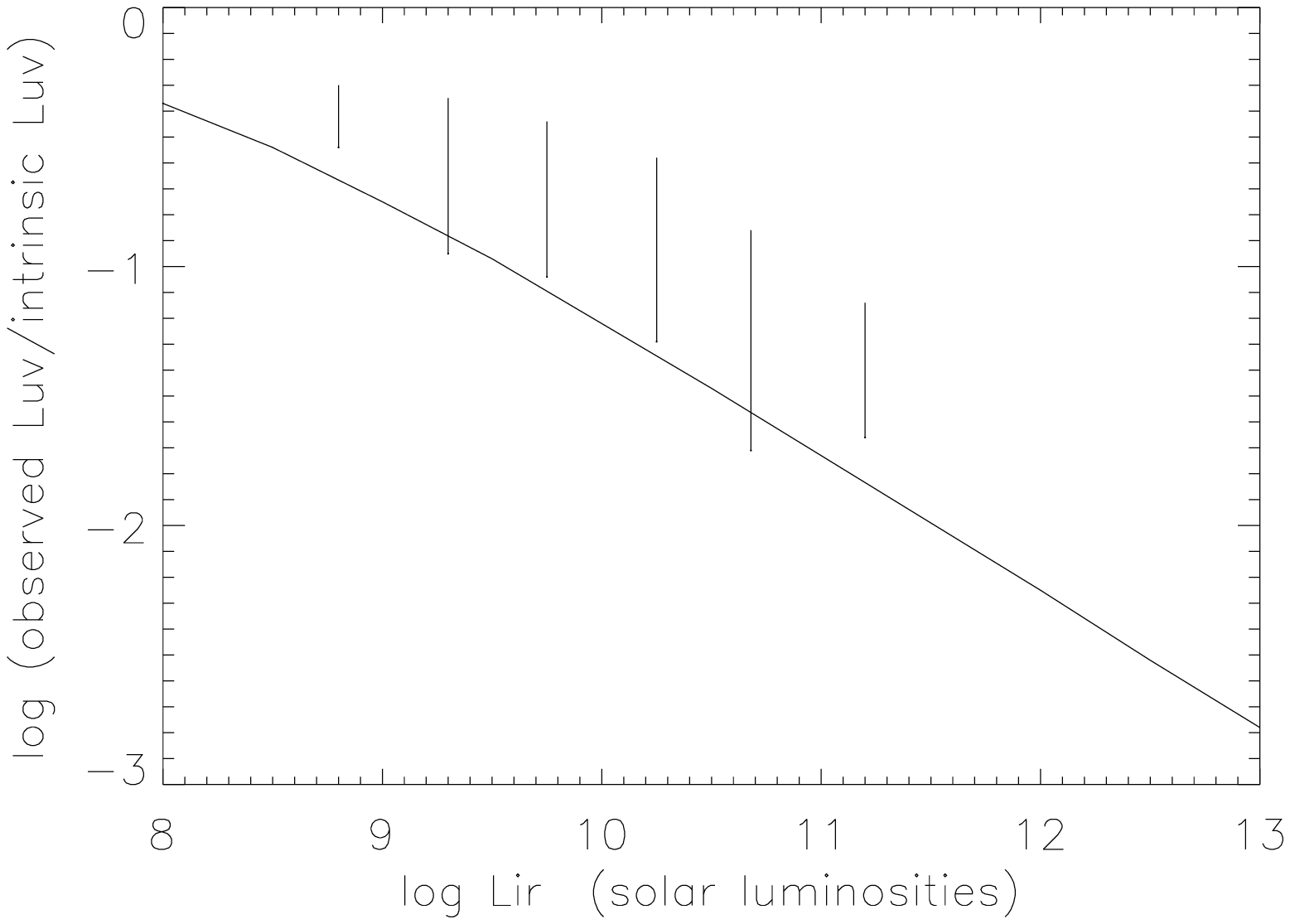}
\caption{Fraction of intrinsic ultraviolet luminosity from starbursts which emerges without being absorbed by dust, defined in text as $L_{uv}$/$L_{uv}$(starburst).  Solid curve shows fraction as measured by SFR(UV)/[SFR(PAH) + SFR(UV)] from equation (2); vertical bars show $\pm$ 1$\sigma$ range of values using $L_{uv}$/($L_{ir}$ + $L_{uv}$) for ultraviolet-selected galaxies in \citet{bua07}. Zero value corresponds to source without any dust extinction.  } 

\end{figure}

\begin{figure}
\figurenum{6}
\includegraphics[scale=0.9]{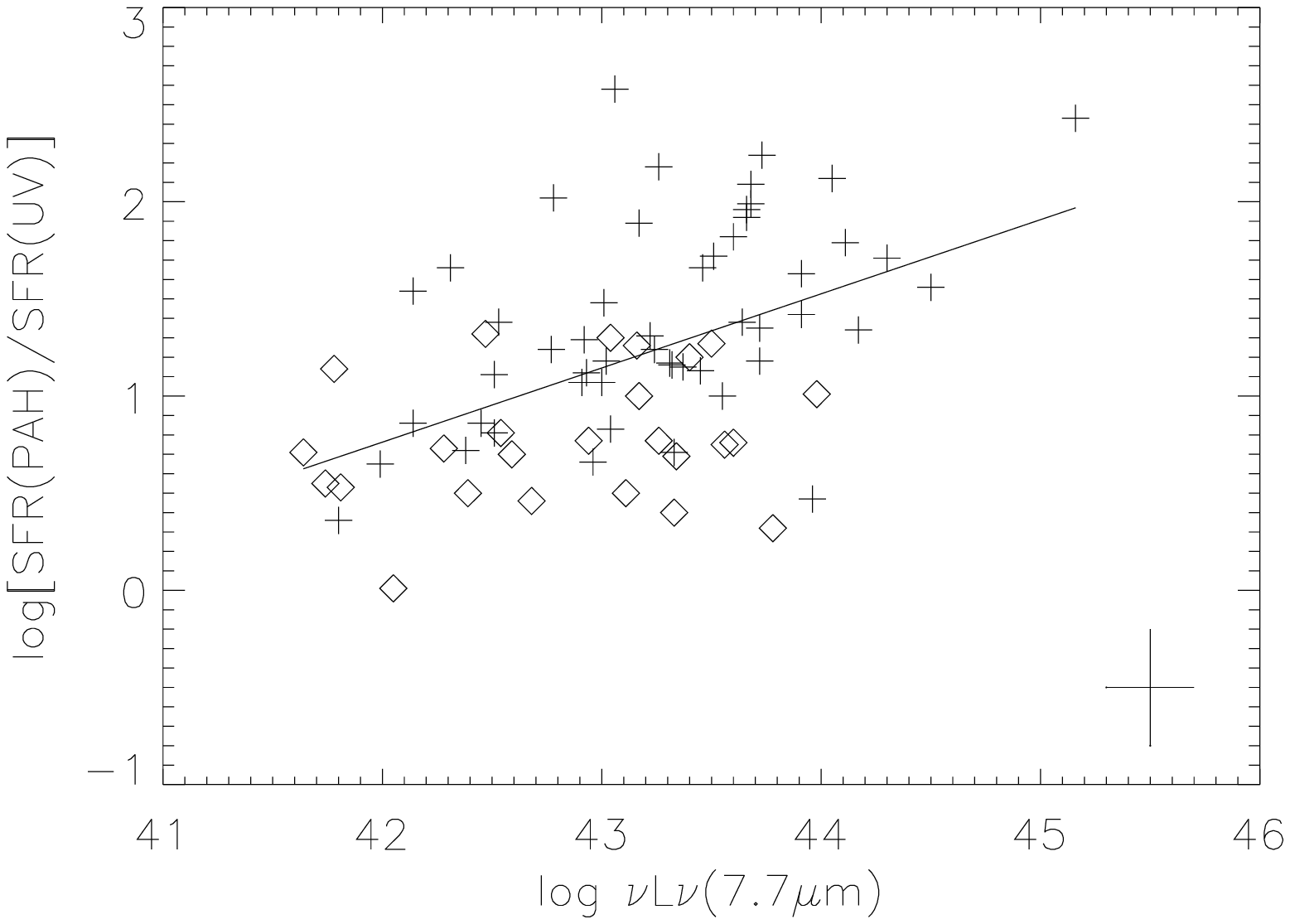}
\caption{Comparison of SFR(UV) measured from ultraviolet continuum luminosity to SFR(PAH) from infrared PAH luminosity for Markarian galaxies in Table 2, using same relations for SFR(UV) and SFR(PAH) as in Figure 4.  Markarian galaxies have f$_{\nu}$(25 $\mu$m) observed by IRAS, which is transformed to f$_{\nu}$(7.7 $\mu$m) using average infrared spectra for sources with 42 $<$ log [$\nu$L$_{\nu}$(15 $\mu$m)] $<$ 44 in \citet{wee09}, for which f$_{\nu}$(25 $\mu$m)/f$_{\nu}$(7.7 $\mu$m) = 2.6$\pm$1.2.  Error cross in lower right hand corner shows 1$\sigma$ uncertainties in both directions; uncertainty for SFR(PAH) arises from uncertainty in determining transformation from $\nu$L$_{\nu}$(7.7 $\mu$m) to the total L$_{ir}$ which determines SFR combined with uncertainty in assumed ratio f$_{\nu}$(25 $\mu$m)/f$_{\nu}$(7.7 $\mu$m); uncertainty in luminosity arises from uncertainty in ratio of f$_{\nu}$(25 $\mu$m)/f$_{\nu}$(7.7 $\mu$m).  Diamonds correspond to upper limits in ratio for sources detected by GALEX but not by IRAS with IRAS limit  taken as 65 mJy; limit is in the sense that SFR(PAH)/SFR(UV) is smaller than limit shown.  Linear fit shown has form log [SFR(PAH)/SFR(UV)]= (0.38$\pm$0.08)log $\nu$L$_{\nu}$(7.7 $\mu$m) - 15.27$\pm$0.25.  (Using empirically determined transformation between $\nu$L$_{\nu}$(7.7 $\mu$m) and total infrared luminosity L$_{ir}$, log L$_{ir}$ = log $\nu$L$_{\nu}$(7.7 $\mu$m) - 32.80$\pm$0.2 for L$_{ir}$ in \ldot~and $\nu$L$_{\nu}$(7.7 $\mu$m) in ergs s$^{-1}$.)}

\end{figure}


\begin{thebibliography}

\bibitem[Ag\"{u}eros et al.(2005)]{ag05}
Ag\"{u}eros, M.A. et al., 2005, \aj, 130, 1022
\bibitem[Balzano(1983)]{bal83}
Balzano, V.A 1983, \apj, 268, 602
\bibitem [Brandl et al.(2006)]{bra06}
Brandl, B. et al. 2006, \apj, 653, 1129.
\bibitem [Calzetti et al.(2007)]{cal07}
Calzetti, D. et al. 2007, \apj, 666, 870
\bibitem[Buat et al.(2007)]{bua07}
Buat, V. et al. 2007, \apjs, 173, 404
\bibitem [Condon(1992)]{con92}
Condon, J. J. 1992, Ann.Rev.Astron.Ap, 30, 575
\bibitem[Condon et al.(2003)]{con03}
Condon, J.J. et al. 2003, \aj, 125, 2411
\bibitem[Dey et al.(2008)]{dey08}
Dey, A. et al. 2008, \apj, 677, 943
\bibitem [Fadda et al.(2006)] {fad06}
Fadda, D. et al. 2006, \aj, 131, 2859. 
\bibitem [Farrah et al.(2008)]{far08}
Farrah, D., et al. 2008, \apj, 677, 957
\bibitem [Genzel et al.(1998)]{gen98}	
Genzel, R. et al. 1998, \apj, 498, 579
\bibitem [Gunn et al.(1998)] {gun98}
Gunn, J. E., et al. 1998, \aj, 116, 3040
\bibitem [Haarsma et al.(2000)] {haa00}
Haarsma, D. B., Partridge, R. B., Windhorst, R. A.,and Richards, E. A. 2000, \apj, 544, 641
\bibitem [Helou et al.(1985)]{hel85}
Helou, G., Soifer, B. T., and Rowan-Robinson, M. 1985, \apj, 298, L7
\bibitem [Higdon et al.(2004)]{hig04}
Higdon, S.J.U., et al. 2004, \pasp, 116, 975
\bibitem [Ho and Keto (2007)]{ho07}
Ho, L, and Keto, E. 2007, \apj, 658, 314
\bibitem [Houck et al.(2004)]{hou04} 
Houck, J. R., et al. 2004, \apjs, 154, 18
\bibitem [Houck et al.(2007)]{hou07} 
Houck, J. R., Weedman, D.W., LeFloc'h, E., and Hao, L. 2007, \apj, 671, 323
\bibitem [Iglesias-Paramo et al.(2006)]{igl06}
Iglesias-Paramo, J. et al. 2006, \apjs, 164, 381
\bibitem [Izotov and Thuan(1998)]{iz98}
Izotov, Y. I., and Thuan, T. X. 1998, \apj, 500, 188
\bibitem [Kennicutt(1998)]{ken98}
Kennicutt, R.C. 1998, Ann.Rev.Astron.Ap, 36, 189
\bibitem [Le Floc'h et al.(2005)]{lef05}
Le Floc'h, E. et al. 2005, \apj, 632, 169
\bibitem[Lonsdale et al(2003)]{lon03}
Lonsdale C. J., et al, 2003, PASP, 115, 897
\bibitem [Madau et al.(1998)]{mad98}
Madau, P., Pozzetti, L., and Dickinson, M. 1998, \apj, 498, 106
\bibitem [Mannucci et al.(2007)]{man07}
Mannucci, F., Buttery, H., Maiolino, R., Marconi, A., and Pozzetti, L. 2007, \aap, 461, 423
\bibitem[Markarian(1967)]{mar67}
Markarian, B.E. 1967, Astrofizika, 3, 55.
\bibitem [Martin et al.(2005)]{mar05}
Martin, D.C. et al. 2005, \apjl, 619, L1.
\bibitem [Martin et al.(2007)]{mar07}
Martin, D.C. et al. 2008, \apjs, 173, 415
\bibitem[Morrissey et al.(2005)]{mor05}
Morrissey, P. et al. 2005, \apjl, 619, L7
\bibitem [Peeters et al.(2004)]{pee04}
Peeters, E., Spoon, H.W.W., and Tielens, A.G.G.M. 2004, \apj, 613, 986
\bibitem [Peeters et al.(2002)]{pee02}
Peeters, E., Hony, S., van Kerckhoven, C., Tielens, A.G.G.M., Allamandola, L.J., Hudgins, D.M., and Bauschlicher,
C.W., 2002, \aap, 390, 1089.
\bibitem[Pope et al.(2008)]{pop08}
Pope, A. et al. 2008, \apj, 675, 1171 
\bibitem [Rigopoulou et al.(2000)]{rig00}	
Rigopoulou, D., Spoon, H. W. W., Genzel, R., Lutz, D., Moorwood, A. F. M., and Tran, Q. D. 2000, \aj, 118, 2625	
\bibitem [Salim et al.(2007)]{sam07}
Salim, S. et al., 2007, \apjs, 173, 267
\bibitem [Smith et al.(2007)]{smi07}
Smith, J.D.T. et al. 2007, \apj, 656, 770
\bibitem[Treyer et al.(2007)]{tre07}
Treyer, M. et al. 2007, \apjs, 173, 256
\bibitem[Weedman et al.(1981)]{wee81}
Weedman, D. W., Feldman, F. R., Balzano, V. A., Ramsey, L. W., Sramek, R. A. and Wu, C.-C., 1981, \apj, 208, 105
\bibitem [Weedman et al.(2006)]{wee06}
Weedman, D.W. et al., 2006, \apj, 653, 101.
\bibitem [Weedman and Houck(2008)]{wee08}
Weedman, D.W., and Houck, J.R. 2008, \apj, 686, 127
\bibitem [Weedman and Houck(2009)]{wee09}
Weedman, D.W., and Houck, J.R. 2009, \apj, 693, 370
\bibitem [White et al.(1997)]{whi97}
White, R. L., Becker, R. H., Helfand, D. J., and Gregg, M. D. 1997, \apj, 475, 479
\bibitem [Wu et al.(2006)]{wu06}
Wu, Y., Charmandaris, V., Hao, L., Brandl, B. R., Bernard-Salas, J., Spoon, H. W. W., and Houck, J. R. 2006, \apj, 639, 157
\bibitem [Yan et al.(2007)]{yan07}
Yan, L. et al. 2007, \apj, 658, 778

\end{thebibliography}
\end{document}